\documentclass[preprint,12pt]{elsarticle}

\usepackage{amssymb}
\usepackage{algorithm}
\usepackage{algpseudocode}
\usepackage{xcolor}

\journal{Optics Communications}

\begin{document}

\begin{frontmatter}



\title{Phase-Space Propagator for Partially Coherent Wave Fields in the Spatial Domain}


\author{Jake J. Rogers, Chanh Q. Tran, Tony Kirk, Paul Di Pasquale, Hong Minh Dao}

\affiliation{organization={Department of Mathematical and Physical Sciences, SCEMS, La Trobe University},
            addressline={Bundoora}, 
            postcode={3086}, 
            state={Victoria},
            country={Australia}}
\affiliation{
    organization={La Trobe Institute of Molecular Science, La Trobe University},
            addressline={Bundoora}, 
            postcode={3086}, 
            state={Victoria},
            country={Australia}}

\author{Pierce Bowman}

\affiliation{organization={School of Physics, The University of Melbourne},
            addressline={Parkville}, 
            postcode={3052}, 
            state={Victoria},
            country={Australia}}

\begin{abstract}
The propagation of wave fields and their interactions with matter are important
for established and emerging fields in optical sciences. Efficient methods for predicting such
behaviour have been employed routinely for coherent sources. However, most real world optical
systems exhibit partial coherence, for which the present mathematical description involves
high dimensional complex functions and hence poses challenges for numerical implementations. 
This demands significant computational resources to determine the properties of partially
coherent wavefields. Here, we describe the novel Phase-Space (PS) propagator, an
efficient and self-consistent technique for free space propagation of wave fields which are partially coherent in the spatial domain. The PS propagator makes use of the fact
that the propagation of a wave field in free space is equivalent to a shearing of the corresponding PSD function.
Computationally, this approach is simpler and the need for using different propagation methods
for near and far-field regions is removed.
\end{abstract}

\begin{graphicalabstract}
\end{graphicalabstract}

\begin{highlights}
\item Description of a novel propagator for partially coherent wave fields in the spatial domain
\item Application of propagation law in free space for optical correlation functions
\item Analysis of computational complexity of different propagation methods for partially coherent wave fields
\end{highlights}

\begin{keyword}
    optical coherence \sep wave field propagator \sep phase space



\end{keyword}

\end{frontmatter}


\section{Introduction}
The propagation of light is an important consideration for any application which involves the interaction between light and matter. In many cases, the extraction of the information of the intercepting sample requires a good understanding of how the wave field propagates from the source plane to the observation plane. A particularly simple subclass of propagation occurs in the case of fully coherent wave fields. The mathematical description of such wave fields and their behaviour is well-established and has been implemented computationally through several efficient algorithms \cite{voelz2009digital,voelz2011computational}. 
\newline
The description of the spatial correlations in quasimonochromatic, statistically stationary wave fields on a two-dimensional (2D) plane involves the use of four-dimensional (4D) functions such as the Mutual Optical Intensity (MOI) function or the Cross Spectral Density (CSD) function \cite{mandel1995optical}. More details will be given in section 2.2.
Numerical methods for propagating partially coherent wave fields are highly computationally demanding, motivating the need for the development of alternative propagation methods. One such approach is that of Coherent Mode Expansion (CME), in which the CSD, or under quasi-monochromatic approximation, MOI, is represented as a superposition of independent coherent modes \cite{wolf1982new}. Each mode may be propagated as a coherent wave field, with their superposition at the observation plane yielding the propagated CSD. This technique converts a 4D problem into a series of 2D problems, which is often more tractable. 
\newline
In practice, the mathematical model used to represent the coherent modes in CME might not correspond to a physical model. Also, the number of modes required to characterise a system is not necessarily small. For wave fields which obey a Gaussian Schell Model (GSM), the number of modes required is inversely proportional to the coherence of the source \cite{starikov1982coherent}. The GSM, in the case of CME, has been used to model important sources such as synchrotrons, in which approximately 300 modes were required to characterise the coherence of the beam (in one dimension) \cite{vartanyants2010coherence}. Similar investigations into fourth-generation sources have found mode counts as high as 90 \cite{khubbutdinov2019coherence} and 200 \cite{glass2017coherent}, with an increasing number of modes required at higher energies. 
\newline
Coherence functions may also be represented in phase space through the PSD function \cite{testorf2009phase}. This allows for free-space propagation of the PSD through a shearing transformation \cite{bastiaans1986application}. Intensity information may also be recovered by a summation along the directional dimension of the PSD \cite{bastiaans1986application}.\newline
In this article, we incorporate the PSD into a full algorithm for the free space propagation of wave fields which are partially coherent in the spatial domain. We refer to this algorithm as the PS propagator which makes use of the unique properties of the PSD and the MOI. 
This technique does not require different algorithms as the propagation distances vary from the near-field to the far-field regions. Additionally, a rigorous analysis of the computational complexity of CME as well as our proposed method will be given. An analysis of the computational complexity presented in Section 4 shows that the PS propagatoris advantagous for reducing computation times for low coherence sources.

\section{Theory and Methodology}
\subsection{Propagation of Quasi-Monochromatic Wave Fields}
The propagation of monochromatic wave fields in free space follows the Helmholtz equation and may be modelled by the well-known Kirchoff-Fresnel \cite{baker1950mathematical} or Rayleigh-Sommerfeld \cite{rayleigh1897xxxvii,sommerfeld1954lectures} theoretical frameworks. Under the paraxial (small-angle) approximation, Fresnel's diffraction formula for quasi-monochromatic wave fields may be written as:
\begin{equation}
	U\left(x,y\right) = \frac{e^{ikz}}{i\lambda z}
	e^{\frac{ik}{2z}\left(x^2+y^2\right)}
	\int_{-\infty}^{\infty}\int_{-\infty}^{\infty}
	\left[U\left(\xi,\eta\right)
	e^{\frac{ik}{2z}\left(\xi^2+\eta^2\right)}
	\right]e^{-\frac{2\pi i}{\lambda z}\left(
		x\xi + y\eta
	\right)} d\xi d\eta
\end{equation}
where $U(x,y)$ and $U(\xi,\eta)$ refer to the propagated wave at the observation plane, and the exit wave at the source plane, respectively. Here also, $\lambda$ refers to the wavelength, $z$ to the propagation distance, and $k$ to the wave number \cite{goodman2005introduction}.
\newline
A number of algorithms exist for the simulation of Fresnel diffraction. Two commonly used methods are the Transfer Function (TF) based propagator and the Impulse Response based (IR) propagator \cite{voelz2009digital,voelz2011computational,goodman2005introduction}. In the case of TF propagation, this distribution is termed the transfer function and has the form:
\begin{equation}
    H\left(k_x, k_y\right) = \exp\left[-i\pi\lambda z \left(k_x^2+k_y^2\right)\right]
\end{equation}
where $k_x$ and $k_y$ denote the Fourier conjugates of $x$ and $y$. This allows the full propagation equation to be expressed:
\begin{equation}
    U_{o}\left(x,y\right) = F^{-1}\{F\{U_{i}\left(x,y\right)\}H\left(k_x,k_y\right)\}
\end{equation}
where $F$ and $F^{-1}$ denote the Fourier and inverse Fourier transforms connecting the input wave field $U_{i}$ to the output wave field $U_{o}$. Similarly, the IR propagator uses the real space factor given by:
\begin{equation}
    h\left(x,y\right) = \frac{e^{ikz}}{i\lambda z}\exp\left[\frac{ik}{2z}\left(x^2+y^2\right)\right]
\end{equation}
so that IR propagation can be written as:
\begin{equation}
    U_{o}\left(x,y\right) = F^{-1}\left\{F\left[U_{i}\left(x,y\right)\right]F\left[h(x,y)\right]\right\}
\end{equation}
Satisfying the oversampling condition is important for ensuring the results of a propagator are accurate. The regions across which the TF and IR propagators are oversampled differ \cite{voelz2009digital,voelz2011computational}. The TF propagator is oversampled when the sample interval $\Delta_{x,y}$ satisfies:
\begin{equation}
    \Delta_{x,y} \geq \frac{\lambda z}{L_{x,y}}
\end{equation}
where $L$ is the side length of the source plane. The IR propagator is oversampled when:
\begin{equation}
    \Delta_{x,y} \leq \frac{\lambda z}{L_{x,y}}
\end{equation}
so that, in principle, these two methods together allow for an accurate calculate of Fresnel diffraction at any propagation distance.

\subsection{Partially Coherent Wave Fields and Coherent Mode Expansion}
For statistically stationary and quasimonochromatic wave fields, the correlation function in the spatial domain can be described as:
\begin{equation}
    J\left(\vec{r}_1,\vec{r}_2\right) = \langle U\left(\vec{r}_1,t\right)U\left(\vec{r}_2,t\right) \rangle
\end{equation}
where the angular brackets indicate the ensemble average of a wavefield $U$ at two different spatial points $\vec{r}_1 \equiv (x_1,y_1)$ and $\vec{r}_2 \equiv (x_2,y_2)$ at the same time, $t$. Here, the statistically stationary condition implies the ensemble avarge $\langle\rangle$ is indepedent of the referent time, $t$, and the quasimonochromatic condition implies that the mean frequency, $\bar{\omega}$ is insignificant compared to the band width $\Delta \omega$, i.e. $\bar{\omega}\Delta \omega \ll 1$. The MOI is related to the CSD by:
\begin{equation}
    W\left(\vec{r}_1,\vec{r}_2,\omega\right) = \int J\left(\vec{r}_1,\vec{r}_2\right)e^{i\bar{\omega}\tau}d\tau
\end{equation}
Previous work \cite{starikov1982coherent} has shown that the CSD can be represented as an orthonormal sum of modes, each of which is fully coherent and obeys the wave equation. Under the quasi-monochromatic approximation, this relation also holds for the MOI. These modes may then be propagated with the TF or IR propagators. Quasi-monochromatic wave fields of an average angular frequency $\omega$ are often represented using the GSM, where the $n$th mode $\phi_n$ for a 1D wave field may be expressed as \cite{starikov1982coherent}:
\begin{equation}
    \phi_n\left(x\right)=\left(\frac{2c}{\pi}\right)^{1/4}\frac{1}{\sqrt{2^n n!}}H_n\left(x\sqrt{2c}\right)e^{-cx^2}
\end{equation}
where $c = \sqrt{a^2 + 2ab}$ and $H_n$ refers to the $n$th order Hermite polynomial. The constants $a$ and $b$ are given by $a = 1/(4\sigma_{I}^2(\omega))$ and $b=1/\left(2\sigma_\mu^2(\omega)\right)$, where $\sigma_{I}(\omega)$ is the intensity waist and $\sigma_\mu(\omega)$ is the coherence length of the beam in the $x$ direction. From which the MOI can be constructed by considering a superposition of such modes:
\begin{equation}
    J\left(x_1, x_2\right) = \sum_n \lambda_n \phi_n^*(x_1)\phi_n(x_2)
\end{equation}
where $\lambda_n$ represents the relative contributions of individual modes following the rule:
\begin{equation}
    \frac{\lambda_n}{\lambda_0} = \left(\frac{b}{a+b+c}\right)^n
\end{equation}
so that the weights of higher order modes can be expressed relative to the zero order weight $\lambda_0$. 
\subsection{The Phase-Space Density Function}
The spatial correlation of a partially coherent wave field on a 2D surface can also be represented by the Phase-Space Density (PSD) function through a change of variables and a Fourier transformation. This is given by:
\begin{equation}
    B(\vec{r},\vec{u}) = \int J'(\vec{r},\vec{\Delta})\exp\left(-i2\pi \vec{u}\cdot\vec{\Delta}\right) d\vec{\Delta}
\end{equation}
where prime denotes the transformed MOI with $\vec{r} = \left(\vec{r}_1+\vec{r}_2\right)/2$ and $\vec{\Delta} = \vec{r}_2 - \vec{r}_1$. The variable $\vec{u}\equiv \left(u_x, u_y\right)$ is then the Fourier conjugate of the separation $\vec{\Delta}$ between the two points $\vec{r}_1$ and $\vec{r}_2$.
Due to the hermitian symmetry of the MOI, the PSD is a real function. However, it can be negative and consequently is often called a quasi-probability distribution function. Essential to this work is a particular geometrical property of the PSD, which may be propagated in free-space through a shearing transformation \cite{bastiaans1986application}, given by:
\begin{equation}
    B_z(\vec{r},\vec{u}) = B_0(\vec{r}-z\vec{u},\vec{u})
\end{equation}
The geometrical property of the PSD described in Equation 13 has been proposed for
recovering the coherence functions of a wave field \cite{raymer1994,nugent1992wave}. Such methods have been demonstrated experimentally for characterising the coherence of synchrotron sources under the separable wave field approximation \cite{tran2005}.
\newline
In this paper, we describe an efficient technique for propagating partially coherent wave fields in the spatial domain by utilising the geometrical property of the PSD. Consider that Equation 8 can be rewritten in terms of $(\vec{r},\vec{\Delta})$ to give:
\begin{equation}
    J\left(\vec{r}_1,\vec{r}_2\right) = \langle U(\vec{r}_1) U(\vec{r}_2) \rangle = \langle A(\vec{r}_1)A(\vec{r}_2)\exp\left(i\left[\phi \left(\vec{r}_1\right)+\phi\left(\vec{r}_2\right)\right]\right) \rangle
\end{equation}
where $A\left(\vec{r}_i\right)$ and $\phi\left(\vec{r}_i\right)$ are the expected values of the amplitude and phase of the wave field at position $\vec{r}_i$. From the substitution of variables required for PSD, this can be expressed as:
\begin{equation}
    J'\left(\vec{r},\vec{\Delta}\right) = \left\langle A\left(\vec{r}+\frac{\vec{\Delta}}{2}\right)A\left(\vec{r}-\frac{\vec{\Delta}}{2}\right)\exp\left[i\phi\left(\vec{r}+\frac{\vec{\Delta}}{2}\right)+i\phi\left(\vec{r}-\frac{\vec{\Delta}}{2}\right)\right]\right\rangle
\end{equation}
For a 1D cross-section of $J'$ where $\vec{\Delta}=2\vec{r}$, this becomes:
\begin{equation}
    J'\left(\vec{r},2\vec{r}\right) = \left\langle A\left(2\vec{r}\right)A(0)\exp\left(i\left[\phi\left(2\vec{r}\right)+\phi\left(0\right)\right]\right)\right\rangle
\end{equation}
which allows for calculating both the amplitude $A(2\vec{r})$ and the phase $\phi(2\vec{r})$ of the wave field relative to the amplitude, $A(\vec{0})$, and phase, $\phi(\vec{0})$, at the central position. 
\newline
This, together with the geometrical property of the PSD as discussed in Equation 13, lays the foundation for a general algorithm for propagating partially coherent wave fields in the spatial domain. Figure 1 provides a visualisation of this algorithm for 2D wave fields represented by $B(\vec{r},\vec{u})$.
Beginning with $J_0\left(\vec{r}_1,\vec{r}_2\right)$ at the source plane as defined by Equation 9 in (a), the change of coordinates produces the transformed MOI denoted by $J_0'\left(\vec{r},\vec{\Delta}\right)$ in (b). A Fourier transform produces the PSD at the source plane in (c) which is equivalent to Equation 12. Propagation occurs by Equation 13 which produces the PSD at the observation plane (d). An inverse Fourier transform along the $\vec{u}$ dimension produces the MOI at the observation plane at a propagation distance $z$ (e) from which the intensity and the phase can be recovered through Equation 16 (f). 
\newline 
\begin{figure}[htbp]
    \centering
    \includegraphics[width=\textwidth]{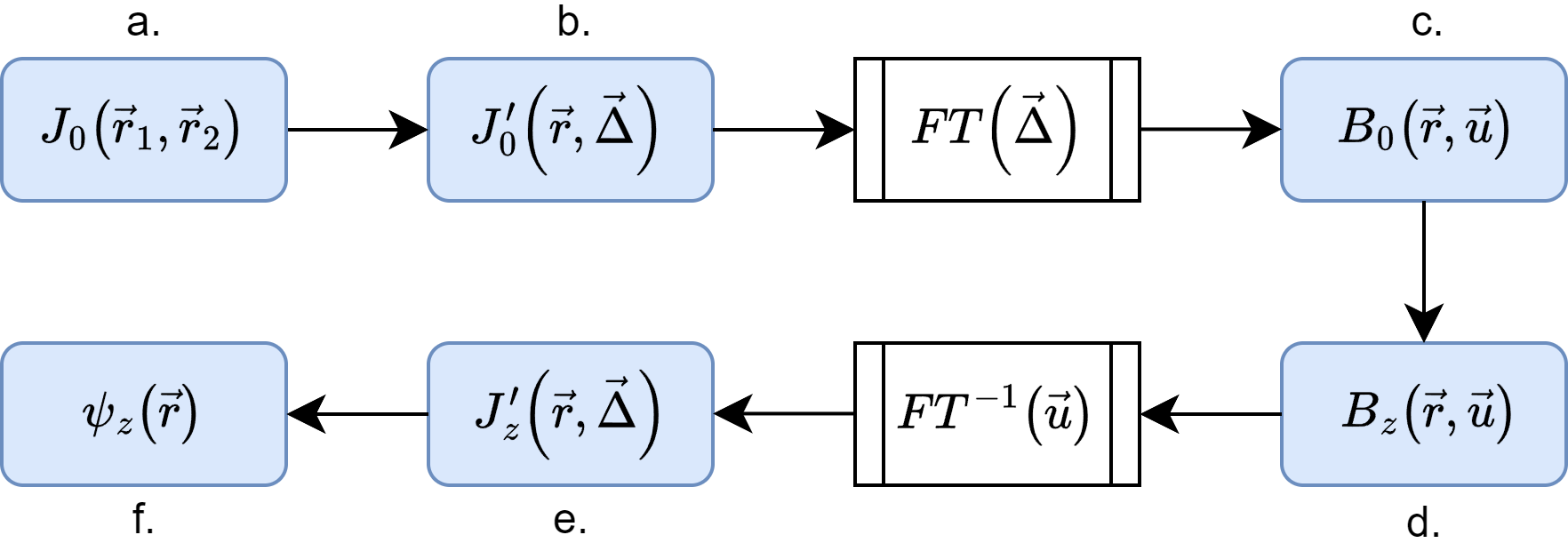}
    \caption{This shows a general algorithm for the propagation of partially coherent wave fields beginning with $J_0\left(\vec{r}_1,\vec{r}_2\right)$ (a). A change of coordinates produces $J'_0\left(\vec{r}, \vec{\Delta}\right)$ (b), where $\vec{r}=(\vec{r}_1+\vec{r}_2)/2$ and $\vec{\Delta} = \vec{r}_2-\vec{r}_1$. A Fourier transformation along the $\vec{\Delta}$ dimension produces the PSD, $B_0\left(\vec{r},\vec{u}\right)$ at the source plane (c) which can be propagated to the observation plane, $B_z\left(\vec{r},\vec{u}\right)$ through a shearing transform (d). The inverse Fourier transform provides the propagated MOI, $J'_z\left(\vec{r},\vec{\Delta}\right)$ (e) from which the propagated intensity and phase can be extracted (f).}
    \label{fig:1}
\end{figure}
\section{Simulation Results}
The proposed algorithm was implemented using the Python programming language with an online open-source repository available \cite{psprop_repo}
\newline
For simplicity, the simulation results presented subsequently are for 1D wave fields only. The corresponding mathematical details can be obtained from Equations 8 - 17 by replacing $\vec{r}=(x,y)$ by $x$ or $y$ and $\vec{\Delta}$ by $\Delta_{x}$ or $\Delta_{y}$. In all subsequent figures, data arrays consisting of $N=1000$ points along one-dimension were used. For 2D distributions, this becomes $1000\times 1000$ points.
\newline
In Figure 2, each distribution detailed in Figure 1 has been simulated and plotted for a plane wave of uniform amplitude passing through a 1D slit of width $a$. The 2D MOI describing the 1D wave field at the source plane $J_0\left(x_1,x_2\right)$ is given in (a), where it was modelled following Equation 14 with $A\left(x_i\right) = 1$ and constant $\phi(x_i)$ for $|x_i|<a$ and $A(X_i)=0$ otherwise, where $a$ is the width of the 1D beam.

The change of coordinates from $x_1$ and $x_2$ to $x$ and $\Delta_x$ as necessary for Equation 12 produces (b), the transformed $J_0'\left(x,\Delta\right)$ which resembles a rotation and dilation of $J_0\left(x_1,x_2\right)$. This change of coordinates was performed by a linear interpolation. The Fourier transform of (b) produces (c) which is the PSD $B_0(x,u_x)$ at the source plane. $B_0\left(x,u_x\right)$ was then propagated to produce $B_z\left(x,u_x\right)$ (d) through a shearing transform which used a first order spline interpolation. The inverse Fourier transform of (d) along the $u_x$ axis provides $J_z'\left(x,\Delta\right)$ in (e). Cross sections of (e) were then used to produce the intensity (top) and phase (bottom) in subplot (f). The phase calculation corresponds to a cross section of (e) taken through the line $(x, 2x)$ which also required linear interpolation. Note that the subfigures are not all represented on the same scale. However, the scale of subfigures a, b, and e is the same, as well as those of c and d are the same, and both plots within subfigure f. 
\begin{figure}[htbp]
    \centering
    \includegraphics[width=\textwidth]{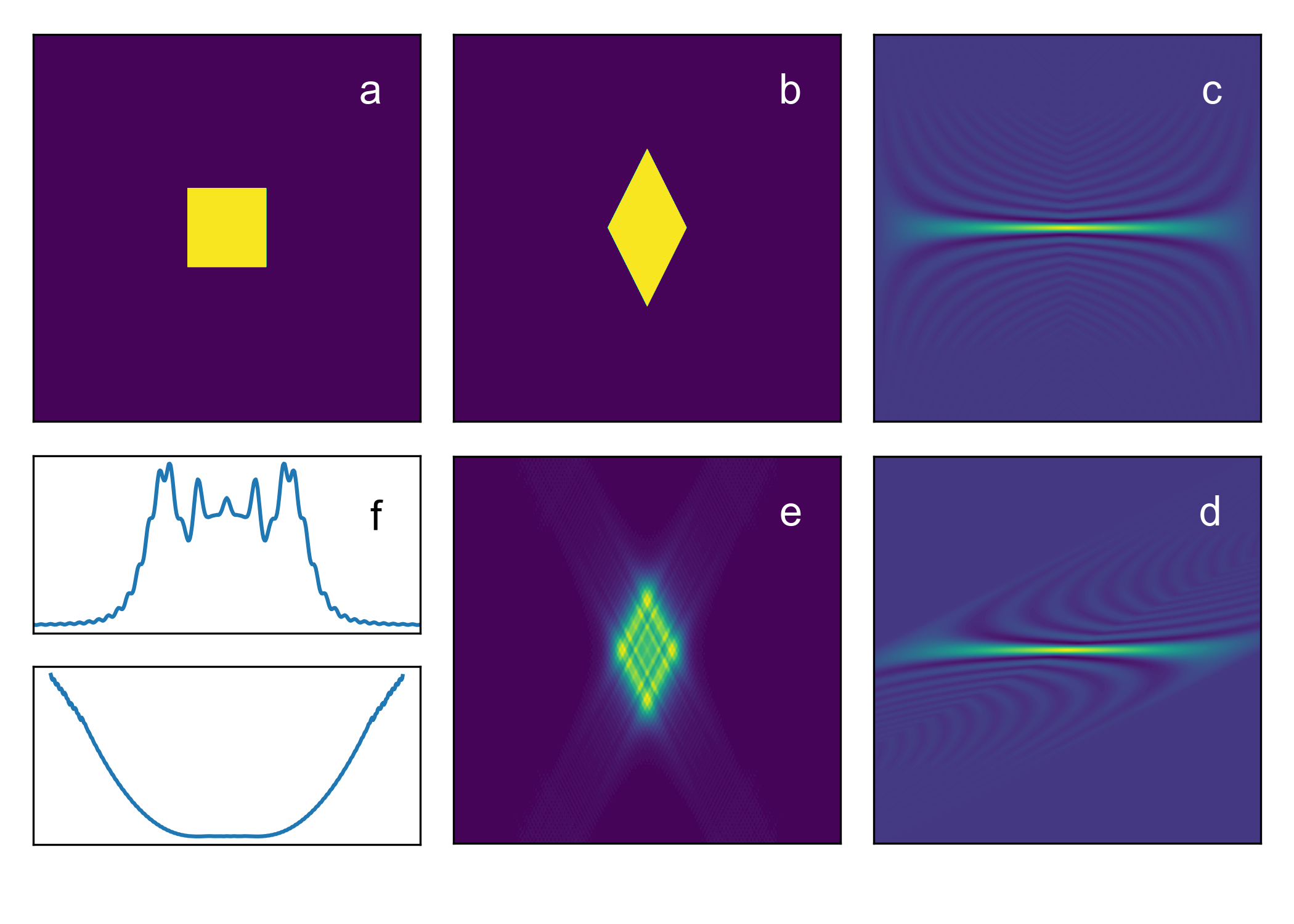}
    \caption{An illustration of the propagation algorithm described in Figure 1 for a 1D wave field. (a) shows $J_0\left(x_1,x_2\right)$ and (b) shows $J_0'(x,\Delta)$ after performing the coordinate change outlined previously, where the subscript $0$ indicates a distribution taken at the source plane. (c) shows $B_0\left(x,u_x\right)$ and (d) shows the effect of propagating the PSD, $B_z\left(x,u_x\right)$, where the subscript $z$ indicates a distribution at the observation plane. Both (c) and (d) are magnified. (e) shows $J_z'\left(x,\Delta\right)$ and (f) shows the intensity and phase profiles calculated after propagation.}
    \label{fig:2}
\end{figure}
In Figure 3, the solid blue lines show the analytical result and the dotted yellow lines show the result predicted by PS propagation. Each column denotes a different Fresnel number, $N_F$, ranging from near to far field. Evidently, the agreement between both methods is excellent except for minor discrepancies in the near field fringes which may be considered a result of numerical error. 
\begin{figure}[htbp]
    \centering
    \includegraphics[width=\textwidth]{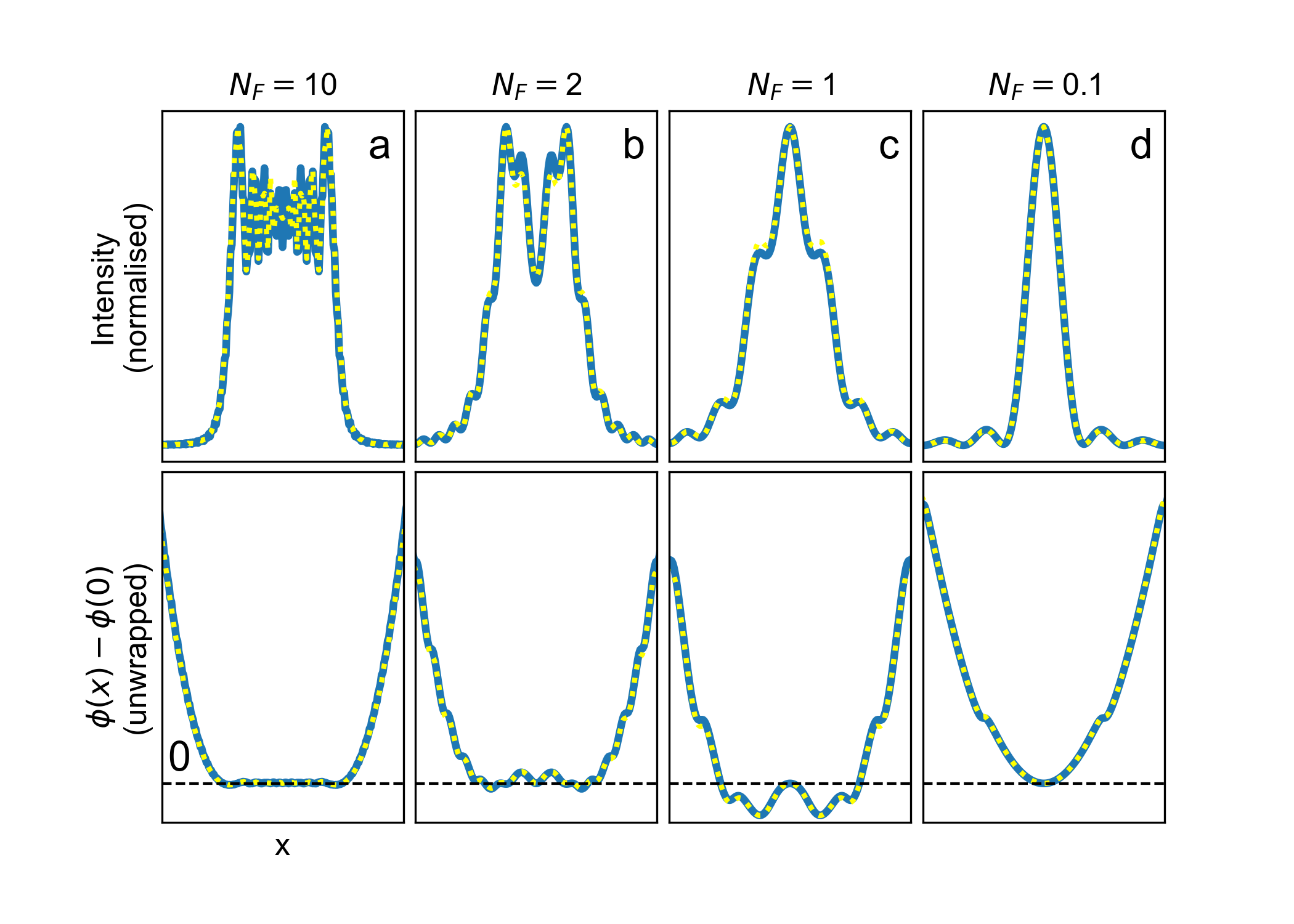}
    \caption{A comparison between the PS propagator (yellow dotted) and the analytical result for a single slit aperture (solid blue). The top row of subplots shows the normalised intensity and the bottom row shows the unwrapped phase. The columns correspond to Fresnel numbers $N_F=$ $10$, $2$, $1$ and $0.1$ from left to right. The excellent agreement between the results obtained using both approaches proves the precision of the PS propagation technique.}
    \label{fig:3}
\end{figure}
~\newline
Figure 4 shows a comparison between the extracted intensity and phase of the PS propagator and conventional propagation of coherent modes using the CME approach. 
The result of CME is given by the solid blue lines and the PS propagator is given by yellow dotted lines. Here, CME was used with the assumption of a GSM source to simulate a partially coherent 1D wave field. The source was modelled after experimental parameters employed in previous research which has modelled a synchrotron beam \cite{flewett2010quantitative,tran2007experimental,mcnulty1996beamline}, with $\sigma_I=471 \times 10^{-6}$  m, $\sigma_\mu = 11.7 \times 10^{-6}$ m, $\lambda = 0.59$ nm. In this simulation, $\sigma_\mu$ was taken to be $20\times 10^{-6}$ m with a single slit of width $100 \times 10^{-6}$ m to increase the visibility of diffraction fringes.
\begin{figure}[htbp]
    \centering
    \includegraphics[width=\textwidth]{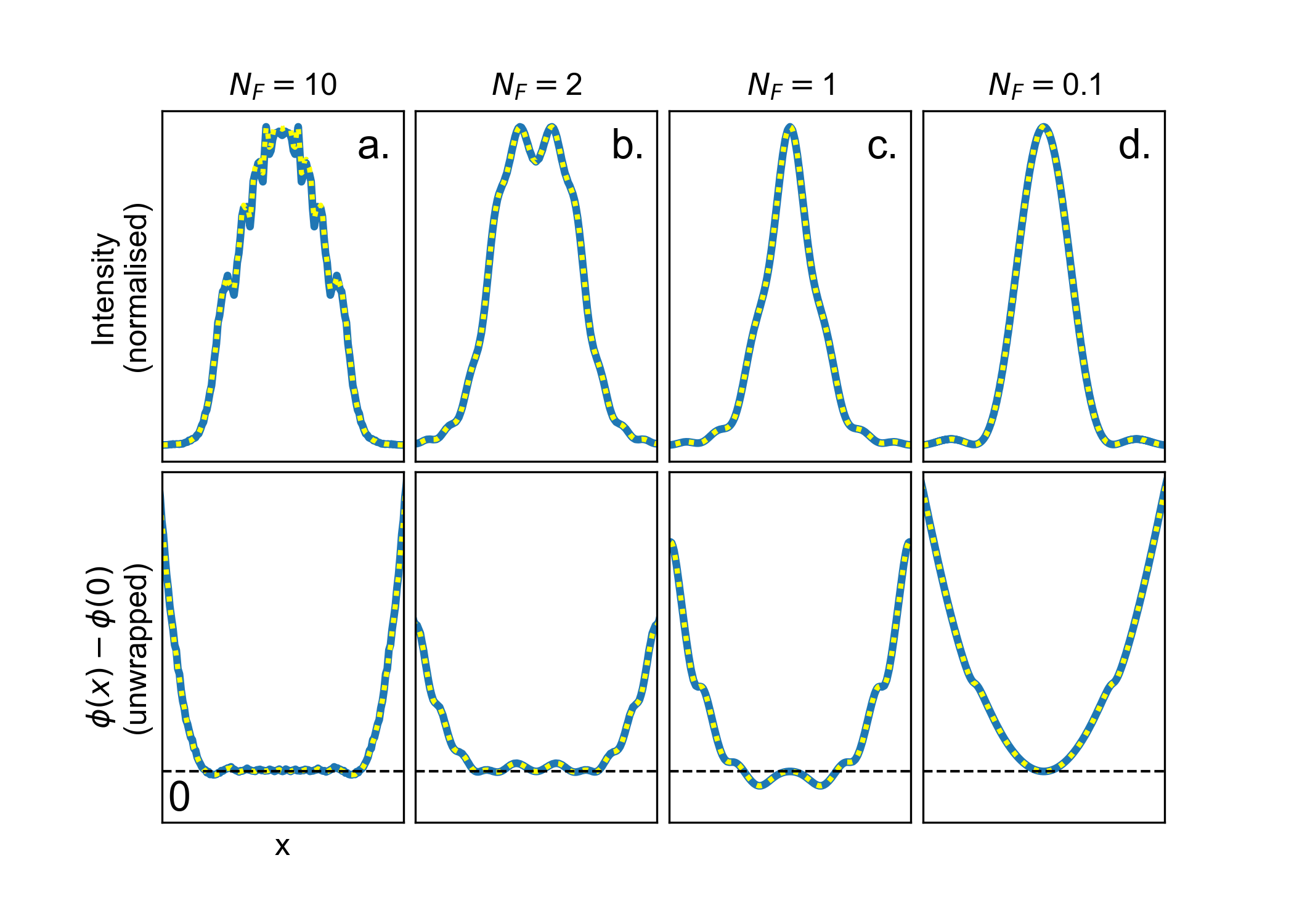}
    \caption{A comparison between the PS propagator (yellow dotted) and the results of CME (solid blue) for propagating a low coherence wave field. The top row of subplots shows the normalised intensity and the bottom row shows the unwrapped phase. The columns show Fresnel numbers $10$, $2$, $1$ and $0.1$ from left to right. }
    \label{fig:4}
\end{figure}
The CME model included 10 modes, each of which was propagated to distances corresponding to different Fresnel numbers. The conventional method chosen was the TF based propagator for $N_F = 10$ and the IR propagator for $N_F = 2, 1, 0.1$. This choice was made based on satisfying the sampling criteria in Equations 6 and 7. At the observation plane, the superposition of these 10 modes was used to produce the propagated MOI from which the wave field was extracted using Equation 17. In the case of the PS propagator, the superposition of these 10 modes was evaluated at the source plane to produce $J_0\left(x_1,x_2\right)$. Propagation and wave field extraction was performed following the method outlined in Figure 1.
\newline
The excellent agreement between these results shown in Figure 4 serves as confirmation of the proposed PS method for partial coherence in the spatial domain. To use the PS propagator, the MOI function at the source plane needs to be predetermined or assumed.
\section{Analysis of Computational Complexity}
The technique of CME requires the propagation of $m$ modes to the observation plane. Commonly used algoriths, such as the TF and IR propagators, perform this propagation through the use of a Fourier transform. 
In the case of a one-dimensional wave field, the fast Fourier transform has computational complexity $O\left(n \log_2 n\right)$ where $n$ refers to the number of data points \cite{cooley1965algorithm}. 
CME which incorporates the fast Fourier transform therefore has a computational efficiency of $O\left( m n \log_2 n \right)$, where $m$ refers to the number of simulated modes.
\newline
Comparatively, the proposed PS propagator requires a one-dimensional Fourier transform and three steps of interpolation. The highest order interpolation used in this simulation was a cubic spline with computational complexity $O(n)$ \cite{press2007numerical}. Evidently, the linearithmic growth of the Fourier transform dominates, so that the PS propagator has a computational efficiency of $O\left(n \log_2 n\right)$. Noteably, there is no $m$ dependence, so that our algorithm's theoretical execution time is equal to the best case ($m = 1$) scenario of CME. From a practical standpoint, the interpolation steps will increase execution time, so that our algorithm is best suited for situations where the number of modes becomes large. 
\section{Conclusion}
A novel technique for the free space propagtion of partially coherent wave fields in the spatial domain has been described and
demonstrated by simulation. This technique, termed the PS propagator, is based on a unique relationship
between the Fourier transform of the PSD function and the propagated wave field. The excellent agreement in results shows that the
method of PS propagation is highly accurate in both the near-field and far-field. Furthermore,
this novel method eliminates the need for different propagation methods at different propagation
distances, which is a common requirement in conventional approaches.
This is a promising technique compared to existing alternatives for modelling partially coherent
wave fields, such as CME where the required number of modes is
highly variable.  The application of the proposed PS propagator to imaging is an exciting alternative prospect. Accurate and
efficient means for determining the amplitude and phase of a wave field are likely to reduce the time and
labour required to perform imaging experiments.
\label{}



\bibliographystyle{elsarticle-num} 
\bibliography{article.bib}

\end{document}